\documentclass[11pt]{article}
\IfFileExists{KLMM/klmm.sty}
  {\usepackage[]{KLMM/klmm}}
  {\usepackage{klmm}}

\newcommand{\wt}{\operatorname{wt}}
\newcommand{\tc}{\operatorname{tc}}
\newcommand{\supp}{\operatorname{supp}}
\newcommand{\leanid}[1]{\nolinkurl{#1}}

\newcommand{\Z}{\mathbb Z}
\newcommand{\Q}{\mathbb Q}
\newcommand{\F}{\mathbb F}

\newcommand{\PP}{\mathsf{PP}}
\newcommand{\BPP}{\mathsf{BPP}}
\newcommand{\PSPACE}{\mathsf{PSPACE}}

\newcommand{\lc}{\operatorname{lc}}
\newcommand{\Disc}{\operatorname{Disc}}
\newcommand{\Li}{\operatorname{Li}}

\newcommand{\norm}[1]{\lVert#1\rVert}
\newcommand{\ind}[1]{\mathbf 1_{\{#1\}}}

\begin{document}
\title{Irreducibility Testing of Sparse Polynomials over Integers is NP-hard and Probabilistic Polynomial Time}
\author{\fontsize{11}{16}\selectfont
    Yichuan Cao\institution{ State Key Laboratory of Mathematical Sciences, Academy of Mathematics and Systems Science, Chinese Academy of Sciences; University of Chinese Academy of Sciences}
    \and Xiao-Shan Gao\instref{1}
    \and Qiao-Long Huang\institution{School of Mathematics, Shandong University}
    \and Jiawei Nie\instref{1}
}
\date{\today}

\maketitle

\begin{abstract}
\noindent
In this paper, we prove that irreducibility testing for sparse polynomials in $\Z[X]$ is coNP-hard under deterministic polynomial-time many-one reductions.
That is, there exists no polynomial-time irreducibility test algorithm unless $\mathrm P=\mathrm{NP}$.
We also provide an upper bound for this problem by demonstrating that the same irreducibility testing problem lies in $\PP\subseteq\PSPACE$; in other words, it can be solved in probabilistic polynomial time and, moreover, within polynomial space.
The fact that $\mathsf{SparseIrred}_{\Z}\in\PP$ relies on the quasi-Riemann hypothesis result of Cureton--Thorner and on the uniform effective prime-ideal bound by Frei--Sofos--Widmer just posted recently.
\end{abstract}

\section{Introduction}
Polynomial factorization is a central problem in algebraic and symbolic computation, and substantial progress has been made in developing efficient algorithms.
This extensive research has produced effective randomized algorithms for polynomial factorization, which have significant applications in complexity theory, coding theory, and cryptography~\cite{ModernCA}.

For dense univariate polynomials, factoring over finite fields admits randomized polynomial-time (Las Vegas) algorithms \cite{CantorZassenhaus1981}, and irreducibility testing (abbr. irreducibility) follows as a consequence.
However, deterministic polynomial-time factorization  over finite fields remains open.
Over integers, factorization and irreducibility are deterministic in polynomial time via the seminal LLL algorithm \citep{LLL1982}.
When polynomials are
instead given by compact representations such as straight-line programs or
sparse (lacunary) sums, the complexity of factorization is surveyed
in~\cite{Kaltofen1989slp}, and a retrospective on the algorithmic progress
in polynomial factorization is given in~\cite{Kaltofen2003success}.

In practical settings and in mathematical software like Maple, polynomials are typically represented in sparse form—that is, as lists of their exponents and coefficients.
Factorization of sparse polynomials is quite different from that of dense ones.
The simple example below illustrates that a reducible sparse polynomial can have an irreducible factor with exponentially many nonzero terms compared to the size of the original polynomial. 
Therefore, there exists no polynomial-time factorization algorithm for sparse polynomials.
\begin{example}
\label{ex-1}
Let $p$ be an odd prime. Then 
\[
f(x)=x^{p}-1  =\prod_{d\mid p}\Phi_d(x)=\Phi_1(x)\Phi_p(x)\in\mathbb{Z}[x]
\]
where each cyclotomic polynomial $\Phi_d(x)$ is irreducible over $\mathbb Z[x]$.
The size of $f(x)$ is  $O(\log p)$.
The polynomial $\Phi_{p}(x)$ contains $p$ nonzero terms, and its size grows exponentially relative to the input size $O(\log p)$ of $f(x)$.
\end{example}
Recently, Mishra \cite{Mishra2026} proves that for every sufficiently large $m$, there exists an $f\in\Q[x]$ with exactly $m$ nonzero coefficients such that \emph{every irreducible factor} of $f$ over $\Q$ has more than $\exp\bigl(c\sqrt{m/\log m}\bigr)$ nonzero coefficients. Thus, for such $f$, every irreducible factor has a superpolynomial number of terms in $m$. 


Therefore, for sparse polynomials, the complexity of testing irreducibility can differ substantially from that of factoring, because a sparse polynomial may have irreducible factors that are dense, making straightforward enumeration of candidate factors infeasible.
Based on the seminal work on the complexity of sparse polynomial operations \citep{Plaisted1984},  Kaltofen and Koiran \citep{KaltofenKoiran2005} proved that irreducibility testing for bi-variate sparse polynomials over integers or sufficiently large finite fields is coNP-hard under randomized polynomial-time many-one reductions.
In \cite{cao2026sparsepolynomialgcdcomputed}, a similar result is proved for arbitrary fields; precisely, 
irreducibility for bi-variate sparse polynomials over any field is coNP-hard under deterministic polynomial-time many-one reductions.

%

In this paper, we consider the following basic problem.

\begin{quote}
\noindent\textbf{Problem $\mathsf{SparseIrred}_{\Z}$.}
\emph{Given a polynomial $f\in\Z[X]$ in sparse form
$\langle f\rangle=\{(e_i,a_i)\}_{i\le\tau}$, of size
$\sum_{i}(\log|a_i|+\log e_i+O(1))$ bits, decide whether $f$ is
irreducible in $\Z[X]$.}
\end{quote}

An instance of the problem is a polynomial $f\in\Z[X]$ given in sparse
(supersparse) form, that is, as a list of its nonzero terms
\begin{equation}
\label{eq-spol}
     \langle f\rangle=\{(e_1,a_1),\ldots,(e_\tau,a_\tau)\},
 \qquad a_i\in\Z\setminus\{0\},\ \ e_i\ \text{pairwise distinct},
\end{equation}

where $\tau=\tc(f)$ is the number of nonzero terms of $f$ and $n$ the degree of $f$. 
Its size is
\[
L= \mathrm{size}(f)=\sum_{i=1}^{\tau}\bigl(\log|a_i|+\log e_i+2\bigr)
 \ \text{bits}.
\]
Only the \emph{logarithms} of the exponents are charged, so a polynomial of
exponentially large degree may still have an input of polynomial length;
consequently, no polynomial-time algorithm may traverse $f$ term by term up
to its degree. The decision problem is to decide, in time polynomial in
$\mathrm{size}(f)$, whether such an $f$ is irreducible in $\Z[X]$
(equivalently, whether $f$ is reducible, i.e.\ $f=gh$ for some nonunits
$g,h\in\Z[X]$). 

The computational complexity of testing irreducibility for univariate sparse polynomials has remained unresolved for a long time, as noted either explicitly or implicitly in \cite{Karpinski1999CH,KaltofenKoiran2005,SparsityChallenges,Roche2018}.
In \cite{Karpinski1999CH}, it was mentioned that ``another interesting related question to which probably can be studied by the method of this paper (Hilbert irreducibility) is deciding irreducibility of sparse polynomials (in $\Z[X]$ ).''
In \cite{KaltofenKoiran2005}, it was mentioned that ``the hardness of sparse irreducibility in $\mathbb{F}_q[X]$ remains open.''
Open problem 7 in \citep{Roche2018} implicitly assumed irreducibility over $\mathbb{Z}$ as an open problem.
Furthermore, the irreducibility of univariate sparse polynomials cannot be deduced from the CoNP-hardness of testing irreducibility for bivariate sparse polynomials because, in \citep{KaltofenKoiran2005}, the polynomials studied have the special shape $F(X,Y) = f(x) + Y^M g(x)$ with $M>0$ and $f\ne0$, and $F(X,Y)$ is irreducible precisely when $f(x)$ and $g(x)$ are coprime.
 
\subsection{Main Result}

In this paper, we present lower and upper bounds for the complexity question of deciding the irreducibility of sparse polynomials over the integers. 
Specifically, we establish the following result (Refer to Theorem \ref{thm:main} and Theorem \ref{thm:main2}).

\begin{theorem}
\label{thm-m1}
$\mathsf{SparseIrred}_{\Z}$ is coNP-hard under deterministic polynomial-time many-one reductions.
Equivalently, reducibility in $\Z[X]$ is NP-hard under the same reductions.
Furthermore, $\mathsf{SparseIrred}_{\Z}\in\PP\subseteq\PSPACE$, that is, 
$\mathsf{SparseIrred}_{\Z}$ can be decided in probabilistic polynomial (PP) time and also in polynomial space ($\PSPACE$).
\end{theorem}
As a consequence of the CoNP-hardness result, a polynomial-time irreducibility test implies $\mathrm P=\mathrm{NP}$.
By Example \ref{ex-1}, the $\PSPACE$ complexity of testing irreducibility is not immediate because the factors of $f$ may have size exponential in that of $f$.

It is worth noting that the proof establishing $\mathsf{SparseIrred}_{\Z}\in\PP$ relies on the recent quasi–Riemann hypothesis result of Cureton--Thorner \cite{CuretonThorner2026} as well as the uniform effective prime-ideal estimate due to Frei--Sofos--Widmer \cite{FreiSofosWidmer2026}.

\begin{remark}
Note that $\PP$ and $\BPP$ are different.
$\BPP$ (Bounded-error Probabilistic Polynomial time) means that 
the algorithm answers correctly with probability $\ge \tfrac{2}{3}$.
$\PP$ means that the algorithm answers correctly with probability $>\tfrac{1}{2}$. 
It is widely believed that  $\BPP \subsetneqq \PP$.
\end{remark}


It should be noted that whether testing irreducibility for univariate sparse polynomials over finite fields is coNP-hard is still an open question.

\subsection{Lean Formalization}
\label{sec:lean}
The accompanying Lean~4 development covers both the hardness results of
Theorem~\ref{thm:main} and the integer upper bound of
Theorem~\ref{thm:main2}, together with their internal mathematical
derivations and machine implementations. The source code, build
instructions, and theorem correspondence are hosted at
\url{https://github.com/MechMath/sparse-irr}.
The project uses Lean \leanid{4.34.0-rc2}, mathlib~\cite{mathlib}, and complexitylib~\cite{complexitylib}.

For the hardness result, the specialized Filaseta--Ford--Konyagin~\cite{FFK}
argument, the sparse reduction, its polynomial running time and output
length, and the one-query oracle reduction are proven without additional
mathematical hypotheses. For the upper bound (Theorem \ref{thm:main2}), the development proves the discriminant and exceptional-prime estimates, the modular root--prime-ideal correspondence, the average estimate, the cutoff and
decision criterion, and the deterministic and strict-majority machine
constructions. The effective prime-ideal counting result obtained from
\cite{CuretonThorner2026,FreiSofosWidmer2026} and the polynomial-time
primality result of \cite{AgrawalKayalSaxena2004} remain explicit theorem hypotheses; the paper's conclusions are derived from them.

All project proofs are free of project-specific axioms and
\leanid{sorry} placeholders; the audited dependency closure uses only
\leanid{propext}, \leanid{Classical.choice}, and \leanid{Quot.sound}.
This audit does not discharge the explicitly stated literature
parameters of the  upper bound. Appendix~\ref{app:lean} gives
the principal declarations, the precise machine specifications, and
reproducibility details.




\section{Irreducibility Test of Sparse Polynomials over Integers is CoNP-hard}
\label{sec-NP}
We will prove the following theorem in this section.

\begin{theorem}
\label{thm:main}
Irreducibility for sparse polynomials in $\Z[X]$ is coNP-hard under deterministic polynomial-time many-one reductions.
Equivalently, reducibility in $\Z[X]$ is NP-hard under the same reductions.
Irreducibility is NP-hard under polynomial-time Turing reductions using one irreducibility query. Consequently, a polynomial-time irreducibility test implies $\mathrm P=\mathrm{NP}$.
\end{theorem}

The theorem is proved by reducing an exact-3-CNF satisfiability formula
$\Phi$ to the sparse polynomial $F=A+X^M(X^N-1)$, where $A$ is
built from the literal polynomials $X^{m_i}-1$ and
$\sum_{r=0}^{q_i-1}X^{rm_i}$ (Refer to Lemma \ref{lem:carrier}) so that the roots of unity of $F$ encode exactly the satisfying assignments.
A gap theorem obtained by specializing the simultaneous-approximation and polynomial-lifting mechanism of Filaseta--Ford--Konyagin \cite{FFK} makes $F$ irreducible precisely when $\Phi$ is unsatisfiable, and every constructed polynomial is presented by a canonical sparse table of polynomial length. 

\subsection{The exact-3-CNF reduction}
The reduction uses exact-3-CNF satisfiability, with repeated literals
permitted. A polynomial with a positive degree is irreducible in $\Z[X]$ if and only if it is primitive and irreducible in $\Q[X]$.
We specify the full reduction, including malformed source strings and the constant/unit convention.

The first reduction step follows \cite[Theorem 5.1]{Plaisted1984}. Note that our construction of polynomial $C_i$ is different from that in \cite[Theorem 5.1]{Plaisted1984}. Also, a new conclusion about $F$ has been added.

\begin{lemma}
\label{lem:carrier}
Let $\Phi$ be a nonempty CNF formula with at least one occurring variable.
Number those variables $0,\ldots,n-1$, and let $q_i$ be the $(i+1)$-st odd
prime. Put $N=\prod_iq_i$ and $m_i=N/q_i$. For a positive or negative literal
on variable $i$ put, respectively,
\[
 P_i^+(X)=X^{m_i}-1,\qquad
 P_i^-(X)=\sum_{r=0}^{q_i-1}X^{rm_i}.
\]
Let $C_j$ be the product of the literal polynomials in clause $j$ and form
the Laurent polynomial
\[
 S(X)=\sum_j C_j(X)C_j(X^{-1}).
\]
Let $d=\max\{|e|:[X^e]S\ne0\}$, $A=3X^dS$, and $B=X^N-1$. Then $S\ne0$,
$A\in\Z[X]$, $A(0)\ne0$, $A^*=A$, where
$A^*=X^{\deg A}A(X^{-1})$, and
\begin{equation}
     \Phi\text{ satisfiable}\quad\Longleftrightarrow\quad
 \gcd_{\Q[X]}(A,B)\ne1. 
 \label{eq:carrier}
\end{equation}
For every integer $M\ge0$, the polynomial $F=A+X^MB$ has a root of unity
if and only if $\Phi$ is satisfiable.
\end{lemma}
\begin{proof}
All $q_i$ and $m_i$ are odd. Hence $P_i^+(-1)=-2$ and
$P_i^-(-1)=\sum_{r=0}^{q_i-1}(-1)^r=1$. No $C_j(-1)$ vanishes, so
$S(-1)=\sum_jC_j(-1)^2>0$. The relation $S(X^{-1})=S(X)$ makes its
coefficients at $-d,d$ equal and nonzero. Consequently, $A$ has degree $2d$,
constant coefficient $3[X^{-d}]S\ne0$, and $A^*=A$.

Let $\zeta^N=1$ and $z_i=\zeta^{m_i}$. Then $z_i^{q_i}=1$.
The polynomial $P_i^+$ vanishes at $\zeta$ exactly when $z_i=1$.
The geometric identity $(z_i-1)P_i^-(\zeta)=z_i^{q_i}-1$ shows that
$P_i^-$ vanishes exactly when $z_i\ne1$: at $z_i=1$ its value is $q_i$.
Declare variable $i$ true when $z_i=1$. A clause is true precisely when
one of its literal polynomials vanishes, hence when $C_j(\zeta)=0$.
Because $|\zeta|=1$ and $C_j$ have real coefficients,
$C_j(\zeta^{-1})=\overline{C_j(\zeta)}$ and
$S(\zeta)=\sum_j|C_j(\zeta)|^2$. Thus, $S(\zeta)=0$ exactly when this
assignment satisfies $\Phi$. Conversely, a chosen truth assignment is
realized by some $\zeta=e^{2\pi it/N}$: by the Chinese remainder theorem
choose $t\equiv0\pmod{q_i}$ for true variables and
$t\equiv1\pmod{q_i}$ for false ones. Since $m_i$ is invertible modulo
$q_i$, the resulting $z_i$ is $1$ exactly in the true case. As
$A(\zeta)=3\zeta^dS(\zeta)$ and $B$ has precisely the $N$-th roots of
unity as roots, this proves \eqref{eq:carrier}.

If $\Phi$ is satisfiable, its witness $\zeta$ satisfies $A(\zeta)=B(\zeta)=0$,
and hence $F(\zeta)=0$. Conversely, suppose $F(\zeta)=0$ for an arbitrary
root of unity. Every conjugate $\eta$ of $\zeta$ is again a root of unity
and is a root of $F$, because $F\in\Z[X]$. For each such $\eta$,
\[
 0\le S(\eta)=\sum_j|C_j(\eta)|^2
   =\frac{|B(\eta)|}{3}\le\frac23.
\]
Here $|\eta|=1$ and $|\eta^N-1|\le2$. The value $S(\zeta)$ is an algebraic
integer: every Laurent monomial in the root of unity $\zeta$ is an algebraic
integer. Its field norm is the product of the displayed conjugate values.
If $S(\zeta)\ne0$, that norm is a nonzero integer, but its absolute value
is at most $(2/3)^{[\Q(\zeta):\Q]}<1$, a contradiction. Thus
$S(\zeta)=0$. The equation $F(\zeta)=0$ now gives $B(\zeta)=0$, so \eqref{eq:carrier}
shows that $\Phi$ is satisfiable.
\end{proof}

\subsection{A Criterion for Irreducibility}

For a nonzero $P=\sum p_eX^e\in\Z[X]$, put
$P^*=X^{\deg P}P(X^{-1})$,
$\wt(P)=\sum_e p_e^2$, and $\tc(P)=|\supp P|$.
The star operation is multiplicative on polynomials with nonzero constant coefficient. The coefficient of $X^{\deg P}$ in $PP^*$
is $\wt(P)$. The polynomial $P^*$ has the same weight and term count
as $P$.
We use the following criterion to test whether a polynomial is irreducible.

\begin{theorem}\label{thm:gap}
Let $A\in\Z[X]$, $N\ge1$, and $E=\deg A$. Assume
\[
 A(0)\ne0,\qquad A^*=A,\qquad \textstyle \gcd_{\Q[X]}(A,X^N-1)=1.
\]
Set $B=X^N-1$, $m=\max(N,E)$, and
$s_0=2(\wt(A)+\tc(A))+8$. If
\begin{equation}
     M\ge(10+2m)64^{s_0},                                  
     \label{eq:gap-bound}
\end{equation}
then, for $F=A+X^MB$ and every $K\in\Z[X]$,
\begin{equation}\label{tag2}
 FF^*=KK^*\quad\Longrightarrow\quad
 K\in\{F,-F,F^*,-F^*\}.  
\end{equation}
If, in addition, $F$ has no complex root of unity, then $F$ is
irreducible in $\Z[X]$.
\end{theorem}

The derivation below specializes the simultaneous-approximation and
polynomial-lift mechanism of Filaseta--Ford--Konyagin \cite{FFK} to this
palindromic family. 

\begin{lemma}
\label{lem:modulus}
Let $L=M+N$ and let $T\subseteq\{0,\ldots,L\}$ have
$1\le s=|T|\le s_0$. Under \eqref{eq:gap-bound}, there are integers
$1\le d\le8^s$ and $\kappa=\lfloor L/d\rfloor$ such that
$\kappa> 2m$, and, with $h=\lfloor\kappa/4\rfloor$, each $t\in T$
has a unique integer $q_t$ with
\begin{equation}\label{tag3}
  0<t+h-q_t\kappa<\kappa/2.    
\end{equation}
In particular, all residues in (\ref{tag3}) are integers strictly between
$0$ and $\kappa/2$.
\end{lemma}
\begin{proof}
Put $R=8^s$. Partition $[0,1)^s$ into $R$ half-open
boxes of side $1/8$. Apply the pigeonhole principle to the $R+1$
points
\[
  \bigl(\{jt/L\}\bigr)_{t\in T},\qquad j=0,\ldots,R.
\]
The difference of two indices gives $1\le d\le R$ and integers
$q_t$ such that
$|dt/L-q_t|<1/8$ for every $t\in T$. Since
$0\le dt/L\le d$, one may take $0\le q_t\le d$.
Write $L/d=\kappa+\delta$, with $0\le\delta<1$. Then
\begin{equation}\label{tag4}
 |t-q_t\kappa|
 \le |t-q_t L/d|+q_t\delta
 < L/(8d)+d
 <(\kappa+1)/8+d.        
\end{equation}
The assumption $L\ge M\ge(10+2m)64^{s_0}\ge
(10+2m)R^2$ gives
\begin{equation}\label{tag5}
 \kappa\ge(10+2m)R-1\ge10R+15,\qquad
 \kappa>2m,\qquad \kappa\ge95.            
\end{equation}
As $d\le R\le(\kappa-15)/10$, (\ref{tag4}) yields
$|t-q_t\kappa|<9\kappa/40-11/8$.
For $\kappa\ge95$, this is less than
$\kappa/4-7/4\le\lfloor\kappa/4\rfloor-1=h-1$.
Consequently
$1<t+h-q_t\kappa<2h-1<\kappa/2$.
Two different multiples of $\kappa$ cannot both give a residue in
$(0,\kappa/2)$, establishing uniqueness.
\end{proof}

\begin{proof}[Proof of Theorem~\ref{thm:gap}]
Put $L=M+N$. Inequality \eqref{eq:gap-bound} gives $M>E$, so $F$ is monic of
degree $L$ and $F(0)=A(0)\ne0$. Palindromicity gives
\begin{equation}\label{tag6}
 F^*=X^{L-E}A-B,\qquad F^*(0)=1.               
\end{equation}
The two terms of $X^MB=X^L-X^M$ do not meet the support of $A$.
Thus
\begin{equation}\label{tag7}
 \wt(F)=\wt(A)+2,\qquad \tc(F)\le\tc(A)+2.      
\end{equation}

Suppose $FF^*=KK^*$. The nonzero constant term of $FF^*$
forces $K(0)\ne0$. Equality of degrees gives $\deg K=L$.
Comparing coefficients of $X^L$ in the products and using (\ref{tag7})
gives
\begin{equation}\label{tag8}
 \wt(K)=\wt(F)=\wt(A)+2,\qquad
 \tc(K)\le\wt(A)+2,      
\end{equation}
because each nonzero integer coefficient has square at least one.
Take
$T=\supp F\cup\supp F^*\cup\supp K\cup\supp K^*$.
By (\ref{tag7})--(\ref{tag8}) and invariance of term count under star,
\begin{equation}\label{tag9}
 |T|\le2(\tc(A)+2)+2(\wt(A)+2)=s_0.              
\end{equation}
The support is nonempty and lies in $[0,L]$.
Apply Lemma~\ref{lem:modulus} to obtain $\kappa,h$.

For any polynomial $P=\sum_{e\in T}p_eX^e$ with support in $T$,
define its shifted lift in $\Z[x,y]$ by
\begin{equation}\label{tag10}
 {\cal L}(P)=\sum_{e\in T}p_e
       x^{r_e}y^{q_e},\qquad
 r_e=e+h-q_e\kappa.      
\end{equation}
Here $q_e=\lfloor(e+h)/\kappa\rfloor$ by (\ref{tag3}), so $q_e\ge0$.
Every $x$-degree $r_e$ is below $\kappa/2$ and
\begin{equation}\label{tag11}
 {\cal L}(P)(x,x^\kappa)=x^hP(x).   
\end{equation}
Every product of two such lifts has $x$-degree strictly below
$\kappa$. On the $\Z$-module spanned by monomials $x^a y^b$
with $0\le a<\kappa$, substitution $y=x^\kappa$ is injective:
distinct pairs $(a,b)$ give distinct integers $a+\kappa b$.
Using (\ref{tag11}) and the equality $FF^*=KK^*$ therefore gives
\begin{equation}\label{tag12}
 {\cal L}(F){\cal L}(F^*)
   ={\cal L}(K){\cal L}(K^*)\quad\text{in }\Z[x,y].       
\end{equation}

We determine the lift of $F^*$. The estimate (\ref{tag5}), with $R\ge8$,
also gives $\kappa\ge16m+79$, hence
$h>m$ and $h+N<\kappa/2$.
Thus the two terms of $-B=1-X^N$ both have $y$-degree zero,
and their lift is $-x^hB(x)$. The support of
$X^{L-E}A$ lies in the interval $[L-E,L]$ of length
$E<\kappa/2$. All its lifted $x$-degrees lie in
$(0,\kappa/2)$ by (\ref{tag3}). If two of its terms had distinct
$y$-degrees, their original exponents would differ by more
than $\kappa/2$, contrary to their difference being at most
$E$. Hence they share one $y$-degree $\ell$. Because
$A(0)\ne0$, the exponent $L-E$ occurs, so for some integer
$\beta\ge0$ the high lift is $x^\beta A(x)y^\ell$.
Also $\ell\ge1$: if $\kappa<L$, then
$L-E+h>\kappa$ since $h>E$; if $\kappa=L$, the same
inequality follows from $h>E$.
Consequently
\begin{equation}\label{tag13}
 {\cal L}(F^*)=x^\beta A(x)y^\ell-x^hB(x),
 \qquad \ell\ge1.        
\end{equation}
Let $c=\min(\beta,h)$, $i=\beta-c$, $j=h-c$.
Then $\min(i,j)=0$, and (\ref{tag13}) equals $x^cG(x,y)$, where
\begin{equation}\label{tga14}   
 G(x,y)=x^iA(x)y^\ell-x^j(X^N-1).             
\end{equation}

We prove that $G$ is a prime element of $\Z[x,y]$.
Since $B(1)=0$ and $\gcd_{\Q[x]}(A,B)=1$, one has
$A(1)\ne0$. In the UFD $\Z[x]$, $x-1$ is prime.
It does not divide the leading coefficient $x^iA(x)$
of $G$ in $y$. It divides every other coefficient:
the intermediate coefficients vanish, and the constant
coefficient is $-x^jB(x)$. Since
\[
 B=(x-1)(1+x+\cdots+x^{N-1}),\qquad
 (1+x+\cdots+x^{N-1})\big|_{x=1}=N\ne0,
\]
$(x-1)^2$ does not divide that constant coefficient.
The two nonzero coefficients $x^iA$ and $x^jB$ are
coprime in $\Z[x]$. Indeed, a nonconstant common
irreducible factor other than $x$ would divide both
$A$ and $B$, contradicting their coprimality over
$\Q[x]$; a common factor $x$ is excluded by
$\min(i,j)=0$, $A(0)\ne0$, and $B(0)=-1$.
A common integer prime is excluded by the coefficient
$-1$ of $B$. Thus $G$ is primitive as a polynomial
in $y$ over $\Z[x]$. Eisenstein's criterion at $x-1$,
followed by Gauss's lemma, makes $G$ irreducible in
$\Z[x,y]$. Since $\Z[x,y]$ is a UFD, $G$ is prime.

From (\ref{tag12})--(\ref{tag13}), $G$ divides
$\mathcal L(K)\mathcal L(K^*)$. Primality gives
\begin{equation}\label{tag15}
     G\mid\mathcal L(K)\quad\text{or}\quad
 G\mid\mathcal L(K^*).   
\end{equation}
Suppose first that $G\mid\mathcal L(K)$. Substitute
$y=x^\kappa$ in this divisibility. Equations (\ref{tag11})
and (\ref{tag13}) give
\begin{equation}\label{tag16}
x^cG(x,x^\kappa)=x^hF^*(x),\qquad
 x^{h-c}F^*(x)\mid x^hK(x).                
\end{equation}
Because $F^*(0)=1$, the polynomials $F^*$ and $x$
are coprime. Euclid's lemma applied to (\ref{tag16}) gives
$F^*\mid K$ in $\Z[x]$. Their degrees are both $L$,
so $K=aF^*$ for an integer $a\ne0$. By (\ref{tag8}),
$\wt(K)=\wt(F^*)=\wt(F)$; hence $a^2=1$ and
$K=\pm F^*$. In the second case of (\ref{tag15}), the same
argument gives $K^*=aF^*$ with $a=\pm1$.
The star operation is involutive on polynomials with
nonzero constant term, so $K=\pm F$. This proves (\ref{tag2}).

Now assume $F$ has no root of unity. From (\ref{tag6}),
\begin{equation}\label{tag17}
 X^{L-E}F-F^*
   =B\bigl(X^{2M+N-E}+1\bigr).           
\end{equation}
The exponent $2M+N-E$ is positive. If $F$ and $F^*$
had a common nonconstant factor over $\Q[X]$, that
factor would have a complex root $z$. Since $F(0)\ne0$,
$z\ne0$. Equation (\ref{tag17}) then gives $z^N=1$ or
$z^{2M+N-E}=-1$. In either case $z$ is a root of
unity, contradicting the assumption. Thus
$\gcd_{\Q[X]}(F,F^*)=1$. Both polynomials are primitive
because $F$ is monic and $F^*(0)=1$, so they have no
nonunit common divisor in $\Z[X]$.

Finally suppose $F=UV$ in $\Z[X]$ with both factors
nonunits. As $F$ is monic, $U,V$ are nonconstant.
Their constant coefficients are nonzero because
$F(0)\ne0$. Put $K=UV^*$. Multiplicativity of star
gives $KK^*=FF^*$. By (\ref{tag2}), $K=\pm F$ or
$K=\pm F^*$. In the first case cancellation of
$U$ from $UV^*=\pm UV$ gives $V^*=\pm V$.
Then $V$ divides both $F=UV$ and
$F^*=U^*V^*$, contrary to coprimality.
In the second case cancellation of $V^*$ from
$UV^*=\pm U^*V^*$ gives $U=\pm U^*$,
and $U$ divides both $F$ and $F^*$, again a
contradiction. Hence no factorization into nonunits
exists; $F$ is irreducible in $\Z[X]$.
\end{proof}


\subsection{Proof of Theorem \ref{thm:main}}
We first present a result about primes.
\begin{lemma}
\label{lem:prime}
Let $L_x=\operatorname{lcm}(1,\ldots,x)$, with $L_0=1$, and let
$\pi(x)$ count the primes at most $x$. With $0^0=1$,
\[
 2^x\le(x+1)L_x\le(x+1)x^{\pi(x)}. \tag{PB1}\label{eq:prime-bound}
\]
If $p_k$ is the zero-indexed $k$-th prime and $q_i=p_{i+1}$, then
$p_k\le64(k+1)^2$ and $q_i\le64(i+2)^2$.
\end{lemma}
\begin{proof}
At $x=0$ the first assertion is $1\le1\le1$. Suppose $x\ge1$.
Fix $0\le j\le x$ and a prime $p\le x$; set
$h_p=\max\{r\ge0:p^r\le x\}$. Unique factorization gives
$v_p(L_x)=h_p$. Counting multiples of $p^r$ in factorials gives
$v_p(n!)=\sum_{r\ge1}\lfloor n/p^r\rfloor$. Hence
\[
 v_p\binom{x}{j}=\sum_{r=1}^{h_p}
 \left(\left\lfloor\frac{x}{p^r}\right\rfloor
 -\left\lfloor\frac{j}{p^r}\right\rfloor
 -\left\lfloor\frac{x-j}{p^r}\right\rfloor\right). \tag{PB2}
\]
Each summand is $0$ or $1$: adding the remainders of $j$ and $x-j$
modulo $p^r$ causes at most one carry. Thus
$v_p\binom{x}{j}\le h_p=v_p(L_x)$.
No prime greater than $x$ divides $\binom{x}{j}$, so
$\binom{x}{j}\mid L_x$. The binomial theorem yields
$2^x=\sum_{j=0}^x\binom{x}{j}\le(x+1)L_x$.
Also $L_x=\prod_{p\le x}p^{h_p}$ and each $p^{h_p}\le x$, which proves
$L_x\le x^{\pi(x)}$.

Fix $k\ge0$, put $t=k+1$ and $X=64t^2$, and suppose
$\pi(X)\le t-1$. The first assertion and $X+1\le2X$ imply
$2^X\le2X^t$. Since $t\le2^t$ for integers $t\ge1$ by induction,
\[
 2X^t=2^{1+6t}t^{2t}
 \le2^{1+6t+2t^2}<2^{64t^2}=2^X.
\]
The strict inequality follows from
$64t^2-(1+6t+2t^2)=62t^2-6t-1\ge56t-1>0$.
This contradiction gives $\pi(X)\ge k+1$, hence $p_k\le X$.
All primes after $2$ are odd, so $q_i=p_{i+1}\le64(i+2)^2$.
\end{proof}

The following lemma provides an estimate for the size of $F$.
\begin{lemma}
\label{lem:size}
For an exact-3-CNF input string of length $b$, the canonical sparse table
of $F=A+X^M(X^N-1)$, with
\begin{equation} \label{eq:w-m-M}
 w=\wt(A)+\tc(A)+4,\qquad
 m=\max(N,\deg A),\qquad
 M=2^{20(w+1)}(m+1)^6, 
\end{equation}
can be computed in deterministic time polynomial in $b$. It has total
binary length polynomial in $b$, and (\ref{eq:w-m-M}) satisfies
\eqref{eq:gap-bound}.
\end{lemma}
\begin{proof}
We use Lemma~\ref{lem:prime}. Its polynomial bound suffices for the
construction.

Only occurring variables are numbered, so $n\le b$. Put
$Q=64(b+2)^2$. Lemma~\ref{lem:prime} bounds all $q_i$ by $Q$. Trial division of
integers at most $Q$ enumerates the first $n$ odd primes in polynomial
time; primality need not be treated as a unit-cost operation. Each
literal polynomial has at most $Q$ terms. An exact-3 clause expands to
at most $Q^3$ raw terms, and $S$ to at most $sQ^6$ raw products, where
$s\le b$ is the number of clauses. Sorting exponents, summing equal
coefficients and deleting zeros take polynomial time in the number and
bit length of these raw terms. A coefficient of $S$ has absolute value
at most $sQ^6$, and its support has at most $sQ^6$ elements. Hence
\[
 \tc(A)\le sQ^6,\qquad \wt(A)\le9(sQ^6)^3,
\]
so the numerical value of $w$ is polynomial in $b$.

The integer $N$ is a product of $n$ numbers at most $Q$, so its binary
length is at most $n\lceil\log_2Q\rceil+1$. Every exponent in a clause
has magnitude at most $3N$, every exponent in $S$ has magnitude at most
$3N$, and $\deg A=2d<6N+1$. The binary length of $m$ is therefore
polynomial in $b$. Binary arithmetic computes (\ref{eq:w-m-M}) with
$O(w+\log(m+1))$ bits. The table of $F$ has no more than
$\tc(A)+2$ entries, and all exponents have at most
$O(w+\log(m+1)+\log N)$ bits. No dense array of length $N$ or $M$ is
formed.

For the numerical inequality, the exponent in \eqref{eq:gap-bound} is $2w$, hence its
power of $64$ is $2^{12w}$. Since $m\ge1$ and $w\ge4$,
\[
 (10+2m)2^{12w}
 \le 2^{20(w+1)}(m+1)^6=M.
\]
For instance, $10+2m\le12(m+1)$ and
$12\le2^{8w+20}(m+1)^5$ prove the displayed comparison directly.
This completes both the algorithmic and threshold assertions.
\end{proof}

\begin{proof}[Proof of Theorem~\ref{thm:main}]
Take a valid nonempty 3-CNF formula with an occurring variable, and
construct $F$ by Lemmas~\ref{lem:carrier} and \ref{lem:size}.
If $\Phi$ is unsatisfiable, \eqref{eq:carrier} makes $A$ and $B$ coprime over $\Q[X]$,
hence they have no nonunit common divisor over $\Z[X]$. Lemma
\ref{lem:carrier} says that $F$ has no root of unity, and
Lemma~\ref{lem:size} supplies \eqref{eq:gap-bound}. Every hypothesis of
Theorem~\ref{thm:gap} is now met, so $F$ is irreducible in $\Z[X]$.

If $\Phi$ is satisfiable, \eqref{eq:carrier} supplies an $N$-th root of unity $\zeta$
with $A(\zeta)=B(\zeta)=0$. Let $\Phi_k$ be the minimal polynomial of
$\zeta$ over $\Q$, where $k$ is its order. The standard cyclotomic
identity says $\Phi_k\in\Z[X]$ is monic, has positive degree, and divides
both $A$ and $B$ in $\Z[X]$: division by the monic $\Phi_k$ is integral,
and a rational remainder vanishing at $\zeta$ must be zero by minimality.
Therefore $\Phi_k\mid F$. Since $k\mid N$, the cyclotomic polynomial
$\Phi_k$ divides $X^N-1$ and has degree at most $N$. The polynomial
$F$ is monic of degree $M+N>N$ because $\deg A<M$; its quotient by
$\Phi_k$ has positive degree. Thus $F$ is a product of two nonunits in
$\Z[X]$.

On malformed 3-CNF strings, output the encoding of $X$. For the empty
formula or a formula with no occurring variable, evaluate its clauses
directly and output $X^2-1$ in the satisfiable case, $X$ otherwise.
These branches are polynomial-time. Consequently this total map sends
3-UNSAT instances exactly to irreducible sparse polynomials, and sends
3-SAT instances exactly to reducible sparse polynomials. The first
equivalence is a many-one reduction from the coNP-complete language
3-UNSAT; the second is a many-one reduction from the NP-complete language
3-SAT. The coding can use the canonical binary sparse representation:
the computed sorted table has nonzero coefficients and distinct exponents,
and a polynomial-time parser rejects noncanonical encodings.

For the Turing claim, a deterministic polynomial-time machine computes
the same $F$, asks an irreducibility oracle about its encoding once, and
negates the answer. This decides 3-SAT. The query is made on a positive
degree polynomial (also in the degenerate branches), so reducible is
exactly the negation of irreducible there. If irreducibility were in
$\mathrm P$, the same computation would put 3-SAT in $\mathrm P$,
and NP-completeness of 3-SAT would give $\mathrm P=\mathrm{NP}$.
\end{proof}

\section{Irreducibility of Sparse Polynomials over Integers is $\PP$}
In this section, we will prove
\begin{theorem}
\label{thm:main2}
Sparse univariate irreducibility over integers satisfies
\[
 \mathsf{SparseIrred}_{\Z}\in\PP\subseteq\PSPACE.
\]
\end{theorem}
Note that both complexities are for deterministic algorithms. 

The proof shows that a fixed zero-free half-plane for the Dedekind zeta functions of all number fields implies $\mathsf{SparseIrred}_{\Z}\in\PP\subseteq\PSPACE$. Applying the zero-free theorem of Cureton--Thorner \cite{CuretonThorner2026} and the uniform effective prime-ideal estimate of Frei--Sofos--Widmer \cite{FreiSofosWidmer2026}, gives the asserted inclusion without assuming GRH. 
The main steps of the proof include: a weighted modular root count detects both distinct factors and repeated factors, an explicit cutoff has $O(L)$ bits, and a deterministic streaming algorithm uses polynomial space. 
A direct majority-machine construction proves the stronger $\PP$ bound.

Let $f$ be a sparse polynomial given in \eqref{eq-spol} with bit size $L$, degree $n$, and $S=\norm f_1$.
Then $\log(n+2)$ and $\log(S+2)$ are $O(L)$. Natural logarithms are denoted by $\log$, binary logarithms by $\log_2$.

We use ring-theoretic irreducibility in $\Z[x]$. 
Thus, a nonconstant input of content greater than one is reducible. For primitive nonconstant inputs, Gauss's lemma identifies irreducibility over $\Z$ and $\Q$. The zero polynomial and units are not irreducible; if constants are allowed, the irreducible constants are the signed rational primes. These cases are handled separately. In the main argument assume only
\begin{equation}\label{eq:normal}
 f\text{ is primitive},\qquad n\ge1.
\end{equation}
In particular, neither squarefreeness nor $f(0)\ne0$ is assumed.

\subsection{Quasi-Riemann hypothesis for Dedekind zeta functions}
The proof uses the recent result on quasi-Riemann hypothesis for Dedekind zeta functions by Cureton--Thorner \cite{CuretonThorner2026} and a result of Frei--Sofos--Widmer \cite{FreiSofosWidmer2026}.

Cureton--Thorner \cite[Theorem~1.1, version~2]{CuretonThorner2026}  prove that, for every number field $K$,
\begin{equation}\label{eq:ct}
 \zeta_K(s)\ne0\qquad\text{if }\Re s>\frac{31}{32},\quad s\ne1.
\end{equation}
The public formalization \cite{AxiomMath2026} contains the statement for an arbitrary type carrying field and number-field structures, with the analytic continuation agreeing with the usual Dedekind zeta function on $\Re s>1$. 

Thus, the unconditional application in this note explicitly relies on \eqref{eq:ct} as an external theorem with a published proof and formalization. If that input is omitted, the reduction below remains a theorem conditional on a uniform positive zero-free gap. It must not be described as a proof from only the older references in the original manuscript.

For a number field $K$, let $d_K=[K:\Q]$, let $D_K$ be its absolute discriminant, and let $\pi_K(x)$ count its nonzero prime ideals of absolute norm at most $x$. Frei--Sofos--Widmer \cite[Theorem~1.1]{FreiSofosWidmer2026} prove that, if $\zeta_K$ has no zeros in $\Re s>1-\delta$, then
\begin{equation}\label{eq:pit}
 |\pi_K(x)-\Li(x)|
 \le Cx^{1-\delta}\bigl(\delta^{-1}\log D_K+d_K\log x\bigr),
 \qquad x>2,
\end{equation}
where $C$ is an effectively computable positive absolute constant. Their theorem is stated for a normal extension $L/K$ and a conjugacy class. Apply it to the identity extension $K/K$. This is normal regardless of whether $K/\Q$ is normal, and gives \eqref{eq:pit} directly. No Galois closure is introduced.

Combining \eqref{eq:ct} and \eqref{eq:pit}, fix once and for all an effectively computable integer $C_0\ge\max\{1,C\}$ such that
\begin{equation}\label{eq:pit32}
 |\pi_K(x)-\Li(x)|
 \le C_0x^{31/32}\bigl(32\log D_K+d_K\log x\bigr)
\end{equation}
for every number field $K$ and every $x>2$. The integer $C_0$ is a fixed machine constant, independent of the input polynomial. No numerical value for it is asserted here; its effective existence is part of the cited analytic estimate and suffices for a uniform complexity-class inclusion.

For $K=\Q$, this yields
\begin{equation}\label{eq:pntQ}
 |\pi(x)-\Li(x)|\le C_0x^{31/32}\log x.
\end{equation}

\subsection{A modular statistic that also detects repeated factors}

For a rational prime $p$, define
\begin{align}
 \nu_f(p)&=\#\{a\in\F_p:f(a)=0\},\label{eq:nu}\\
 \mu_f(p)&=\#\{a\in\F_p:f(a)=f'(a)=0\},\label{eq:mu}\\
 A_f(X)&=\sum_{p\le X}\bigl(\nu_f(p)+\mu_f(p)\bigr),
 &P(X)&=\pi(X).\label{eq:counts}
\end{align}
Because $f$ is primitive, its reduction is never the zero polynomial. Hence
\begin{equation}\label{eq:rootbound}
 0\le\mu_f(p)\le\nu_f(p)\le n
\end{equation}
even at primes where the degree drops or the reduction is not squarefree.

Write the rational factorization, used only in the proof, as
\begin{equation}\label{eq:factor}
 f=\varepsilon\prod_{i=1}^{r}h_i^{m_i},\qquad
 \varepsilon\in\{1,-1\},\quad m_i\ge1,
\end{equation}
with distinct primitive irreducible $h_i\in\Z[x]$ of positive leading coefficient. Put
\begin{equation}\label{eq:weights}
 w_i=1+\ind{m_i\ge2},\qquad
 \rho(f)=\sum_{i=1}^{r}w_i
 =r+\#\{i:m_i\ge2\}.
\end{equation}
Since $w_i\le m_i$,
\begin{equation}\label{eq:weighteddegree}
 \sum_iw_i\deg h_i\le n,\qquad 1\le\rho(f)\le n.
\end{equation}
Most importantly,
\begin{equation}\label{eq:rho}
 \boxed{f\text{ irreducible over }\Q\iff\rho(f)=1.}
\end{equation}
For example, a power $h^m$ with irreducible $h$ and $m\ge2$ has $\rho(f)=2$, not one. This is why counting only distinct modular roots would not suffice without a squarefreeness test.

Let
\[
 F=\prod_i h_i,\qquad N=\deg F\le n,
 \qquad Z=\lc(F)\Disc(F)\ne0.
\]
The polynomial $F$ and the integer $Z$ will never be computed. They identify a finite exceptional set for the proof. If $p\nmid Z$, the reductions of the $h_i$ are pairwise coprime and separable, with unchanged degrees. At a root $a$ of $\bar h_i$,
\[
 \bar f'(a)\ne0\iff m_i=1.
\]
Indeed, when $m_i=1$, the derivative at $a$ is the product of the nonzero value $\bar h_i'(a)$ and the nonzero values of all other factors. When $m_i\ge2$, every derivative term retains a factor $\bar h_i$. Therefore
\begin{equation}\label{eq:localidentity}
 \nu_f(p)+\mu_f(p)=\sum_iw_i\nu_{h_i}(p)
 \qquad(p\nmid Z).
\end{equation}
No exclusion of primes dividing the multiplicities $m_i$ is needed.

\subsection{Uniform discriminant and exceptional-prime bounds}

Set
\begin{equation}\label{eq:Delta}
 u=\lceil\log_2(n+2)\rceil,\quad
 v=\lceil\log_2(S+2)\rceil,\quad
 \Delta=u+v+2.
\end{equation}
In particular, $\Delta\ge6$ and $\Delta=O(L)$.
We have the following discriminant bounds.
\begin{lemma}
\label{lem:disc}
Let $K_i=\Q(\alpha_i)$, where $\alpha_i$ is a root of $h_i$, and let $D_i$ be the absolute discriminant of $K_i$. Then
\begin{equation}\label{eq:Dsum}
 \sum_iw_i\log D_i\le n\log n+n(n-1)\log S\le n^2\Delta.
\end{equation}
Furthermore, if $b$ is the number of distinct rational primes dividing $Z$, then
\begin{equation}\label{eq:badcount}
 b\le2n\Delta.
\end{equation}
\end{lemma}
\begin{proof}
For an integer polynomial $h=a\prod_{j=1}^{d}(x-\alpha_j)$, let
\[
 M(h)=|a|\prod_{j=1}^{d}\max\{1,|\alpha_j|\}
\]
be its Mahler measure. Its multiplicativity and the inequality $M(f)\le\norm f_1=S$ imply
\[
 M(h_i)\le S,\qquad M(F)\le S,
\]
because every nonzero integer polynomial has Mahler measure at least one.

For a separable polynomial $h$ of degree $d$, Hadamard's inequality applied to its Vandermonde determinant gives
\begin{equation}\label{eq:mahlerdisc}
 |\Disc(h)|\le d^dM(h)^{2d-2}.
\end{equation}
In detail, the row corresponding to $\alpha_j$ has Euclidean norm at most $\sqrt d\max\{1,|\alpha_j|\}^{d-1}$; squaring the determinant bound and multiplying by $|a|^{2d-2}$ gives the assertion. For degree one the discriminant is one, consistently with this formula.

To handle nonmonic factors without any implicit integrality assumption, put $a_i=\lc(h_i)$, $d_i=\deg h_i$, and
\[
 g_i(y)=a_i^{d_i-1}h_i(y/a_i)\in\Z[y].
\]
This is a monic irreducible polynomial for $a_i\alpha_i$, and
\[
 \Disc(g_i)=a_i^{(d_i-1)(d_i-2)}\Disc(h_i).
\]
The field discriminant divides the discriminant of the order $\Z[a_i\alpha_i]$, so
\[
 \log D_i\le d_i\log d_i+d_i(d_i-1)\log S,
\]
using $a_i\le M(h_i)\le S$ and \eqref{eq:mahlerdisc}. Multiplying by $w_i$ and using \eqref{eq:weighteddegree} proves
\[
 \sum_iw_i\log D_i
 \le n\log n+n(n-1)\log S\le n^2\Delta.
\]

Similarly, $\lc(F)\le M(F)\le S$ and \eqref{eq:mahlerdisc} give
\[
 \log_2|Z|\le N\log_2N+(2N-1)\log_2S
 \le2n\Delta.
\]
A nonzero integer has at most $\log_2$ of its absolute value distinct prime divisors; if $|Z|=1$, there are none. This proves \eqref{eq:badcount}.
\end{proof}

\begin{lemma}[Degree-one prime ideals]\label{lem:dedekind}
For $p\nmid Z$, the number $\nu_{h_i}(p)$ equals the number of prime ideals of $K_i$ above $p$ with residue degree one.
\end{lemma}
\begin{proof}
The prime $p$ divides neither $a_i$ nor $\Disc(h_i)$, and consequently does not divide $\Disc(g_i)$. It therefore divides neither the field discriminant nor the index of $\Z[a_i\alpha_i]$ in the full ring of integers. Dedekind's factorization theorem applies to $g_i$; its distinct irreducible factors modulo $p$ record the residue degrees. The invertible change of variable $y=a_ix$ identifies the roots of $g_i$ and $h_i$ modulo $p$. See, for example, \cite[Chapter~3]{MilneANT} for the factorization theorem and its index condition.
\end{proof}

\subsection{The complete prime-root average estimate}
We have the following uniform error estimate.
\begin{proposition}
\label{prop:average}
For every $X>2$,
\begin{equation}\label{eq:average}
 \begin{split}
 |A_f(X)-\rho(f)\pi(X)|\le{}&
 C_0X^{31/32}\bigl(64n^2\Delta+4n\log X\bigr)\\
 &+2n\sqrt X+4n^2\Delta.
 \end{split}
\end{equation}
\end{proposition}
\begin{proof}
Let $\pi_{K_i}^{(1)}(X)$ count prime ideals of $K_i$ with norm at most $X$ and residue degree one over $\Q$. A prime ideal with higher residue degree has norm $p^a\le X$ with $a\ge2$, hence $p\le\sqrt X$. There are at most $d_i$ prime ideals above a fixed rational prime. Therefore
\[
 0\le\sum_iw_i\bigl(\pi_{K_i}(X)-\pi_{K_i}^{(1)}(X)\bigr)
 \le n\sqrt X\le2n\sqrt X.
\]
At each $p\nmid Z$, \eqref{eq:localidentity} and Lemma~\ref{lem:dedekind} identify the contribution to $A_f(X)$ with the corresponding weighted degree-one prime-ideal contribution. At each exceptional prime, both contributions lie in $[0,2n]$, so their difference in absolute value is at most $2n$. Lemma~\ref{lem:disc} gives
\begin{equation}\label{eq:baderror}
 \left|A_f(X)-\sum_iw_i\pi_{K_i}(X)\right|
 \le2n\sqrt X+4n^2\Delta.
\end{equation}
Applying \eqref{eq:pit32} to every $K_i$ and adding with weights gives
\[
 \left|\sum_iw_i\pi_{K_i}(X)-\rho(f)\Li(X)\right|
 \le C_0X^{31/32}
 \left(32\sum_iw_i\log D_i+\sum_iw_id_i\log X\right).
\]
By \eqref{eq:pntQ} and $\rho(f)\le n$, the replacement of $\rho(f)\Li(X)$ by $\rho(f)\pi(X)$ costs at most
\[
 C_0nX^{31/32}\log X.
\]
Combining these estimates with \eqref{eq:Dsum} proves the stronger bound with $32n^2\Delta+2n\log X$ in the bracket. The stated, deliberately looser bound \eqref{eq:average} follows.
\end{proof}

\subsection{An explicit cutoff with linearly many bits}

The following formulas use only integers of $O(L)$ bits:
\begin{equation}\label{eq:cutoff}
 U=2^{20}C_0n^2(\Delta+1)^2,\qquad
 \ell=\lceil\log_2 U\rceil,\qquad
 k=4\ell,\qquad X=2^{32k}.
\end{equation}
The integer $\ell$ is computed by a bit-length operation, with the usual adjustment when $U$ is a power of two. The algorithm need not evaluate logarithms of real numbers.

\begin{lemma}\label{lem:cutoff}
The cutoff in \eqref{eq:cutoff} satisfies
\begin{equation}\label{eq:cutoffgap}
 2^k\ge2^{20}C_0n^2(\Delta+k)^2,
\end{equation}
\begin{equation}\label{eq:Plower}
 \pi(X)\ge\frac{X}{4\log X}\ge\frac{X}{128k},
\end{equation}
and
\begin{equation}\label{eq:rounding}
 \left|\frac{A_f(X)}{\pi(X)}-\rho(f)\right|<\frac14.
\end{equation}
Also, $\log_2X=32k=O(L)$.
\end{lemma}
\begin{proof}
We have $\ell\ge20$ and $2^\ell\ge U$. Since
\[
 \Delta+4\ell\le4\ell(\Delta+1),\qquad 2^{3\ell}\ge16\ell^2\quad(\ell\ge2),
\]
we get
\[
 \begin{split}
 2^{20}C_0n^2(\Delta+k)^2
 &\le2^\ell\left(\frac{\Delta+4\ell}{\Delta+1}\right)^2\\
 &\le2^\ell16\ell^2\le2^{4\ell}=2^k,
 \end{split}
\]
which proves \eqref{eq:cutoffgap}.

For $X\ge4$,
\[
 \Li(X)\ge\int_{X/2}^{X}\frac{dt}{\log X}=\frac{X}{2\log X}.
\]
Since $\log X\le32k$, \eqref{eq:cutoffgap} implies
\[
 2^k\ge4C_0(\log X)^2.
\]
As $X^{1/32}=2^k$, \eqref{eq:pntQ} yields \eqref{eq:Plower}.

Using $\log X\le32k$, the first term in \eqref{eq:average} is bounded by
\[
 128C_0n^2(\Delta+k)\frac{X}{2^k}.
\]
Divide \eqref{eq:average} by \eqref{eq:Plower}. The resulting error is at most
\begin{equation}\label{eq:three}
 \frac{2^{14}C_0n^2k(\Delta+k)}{2^k}
 +\frac{256kn}{2^{16k}}
 +\frac{512kn^2\Delta}{2^{32k}}.
\end{equation}
The first term is at most $1/64$ by \eqref{eq:cutoffgap}. For the second, the same inequality implies
\[
 kn\le2^{k/2-10},
\]
so that term is at most $2^{-2-31k/2}<1/64$. For the third, use
\[
 n^2\le\frac{2^k}{2^{20}C_0(\Delta+k)^2},\qquad
 \frac{k\Delta}{(\Delta+k)^2}\le\frac14.
\]
It is at most $2^{-13-31k}<1/64$. Thus \eqref{eq:three} is less than $3/64<1/4$.

Finally,
\[
 \ell\le21+\log_2C_0+2\log_2 n+2\log_2(\Delta+1)=O(L).
\]
The constant $C_0$ is fixed, so $k$ and the bit length of $X$ have the claimed bounds.
\end{proof}

\begin{corollary}
\label{cor:decision}
For a primitive nonconstant input and the cutoff \eqref{eq:cutoff},
\begin{equation}\label{eq:decision}
 {f\text{ irreducible over }\Q\iff 2A_f(X)<3\pi(X).}
\end{equation}
\end{corollary}
\begin{proof}
If $f$ is irreducible, then $\rho(f)=1$ and $A_f(X)/\pi(X)<5/4<3/2$. If $f$ is reducible, then $\rho(f)\ge2$ and $A_f(X)/\pi(X)>7/4>3/2$. Apply Lemma~\ref{lem:cutoff} and \eqref{eq:rho}.
\end{proof}

\subsection{The deterministic polynomial-space algorithm and $\mathsf{SparseIrred}_{\Z}\in\PSPACE$}

The main loop is the following. Integers $p<X$ suffice because $X$ is a power of two greater than two and hence is not prime.
\begin{center}
\begin{minipage}{0.93\textwidth}
\begin{verbatim}
Normalize the sparse input.
Handle zero, units, constants, and nonprimitive inputs.
Compute n, S, Delta, U, ell, k, and X from the formulas above.
P := 0; A := 0.
For p := 2, ..., X-1:
    If p is prime:
        P := P + 1.
        For a := 0, ..., p-1:
            If f(a) = 0 modulo p:
                A := A + 1.
                If f'(a) = 0 modulo p:
                    A := A + 1.
Accept exactly when 2*A < 3*P.
\end{verbatim}
\end{minipage}
\end{center}

Primality in this deterministic implementation can be tested by trial division. Neither a fast primality algorithm nor a factorization oracle is needed for the space bound. For modular evaluation, use
\[
 f(a)\equiv\sum_j a_ja^{e_j}\pmod p,
\qquad
 f'(a)\equiv\sum_{j:e_j>0}a_je_ja^{e_j-1}\pmod p.
\]
Binary modular exponentiation takes space polynomial in the lengths of $p$, the coefficient, and the exponent. Here each is $O(L)$; derivative coefficients also have $O(L)$ bits. The exponent-zero convention evaluates constant terms correctly, including at $a=0$.

The integer $X$ has $O(L)$ bits. Thus the prime counter, trial divisor, field-element counter, and modular arithmetic registers all have $O(L)$ bits. By \eqref{eq:rootbound},
\[
 0<P\le X,\qquad 0\le A\le2nP\le2nX.
\]
The two accumulators therefore have $O(L)$ bits as well. The input normalization and content calculation are polynomial-time and polynomial-space; their workspace is reused. Every loop is finite, and no vector of length $n$, no prime list, no polynomial factor, and no dense squarefree part is stored.

Corollary~\ref{cor:decision} proves correctness. This already proves the requested $\PSPACE$ inclusion. The core enumeration after normalization, in fact, admits $O(L)$ work space with ordinary binary arithmetic; a linear-space preprocessing claim is not needed here.

\subsection{A stronger bound: $\mathsf{SparseIrred}_{\Z}\in\PP$}

The same decision inequality admits a polynomial-time majority implementation. This section is not needed for the deterministic space algorithm but confirms that the result does not conceal a long witness-consistency problem.

Let $m=32k$, so $X=2^m$ and $m=O(L)$. Define polynomial-time predicates on a prime candidate $p$ and residue candidate $a$, each encoded in exactly $m$ bits, and a bit $b$:
\begin{align*}
 R_+(p,a,b)&\iff p\text{ prime},\ a=0,\ b=0,\\
 R_-(p,a,b)&\iff p\text{ prime},\ a<p,\ f(a)=0\pmod p,\\
 &\hspace{36mm}\text{and }\bigl(b=0\ \text{or}\ f'(a)=0\pmod p\bigr).
\end{align*}
Their satisfying-assignment counts are respectively $P=\pi(X)$ and $A=A_f(X)$. Primality is decidable in polynomial time by Agrawal--Kayal--Saxena \cite{AgrawalKayalSaxena2004}; the sparse modular evaluations are polynomial time in $L+m=O(L)$. The other preprocessing steps are also polynomial time.

For completeness, construct a majority machine explicitly. Guess three mode bits, then $(p,a,b)$, then one additional fair bit $c$. For modes $0,1,2$, accept regardless of $c$ when $R_+$ holds; otherwise, accept exactly when $c=1$. For modes $3,4$, reject regardless of $c$ when $R_-$ holds; otherwise, accept exactly when $c=1$. For modes $5,6,7$, accept exactly when $c=1$.

Every branch uses polynomial time and the same number of guesses. In the difference between accepting and rejecting paths, each satisfying assignment for each positive mode contributes $2$, each satisfying assignment for each negative mode contributes $-2$, and all remaining assignments contribute zero. Consequently,
\[
 \#\mathrm{accept}-\#\mathrm{reject}=2(3P-2A).
\]
The machine has a strict accepting majority if and only if \eqref{eq:decision} holds. Deterministic preprocessing cases can immediately accept or reject, padding with unused fair bits if desired. This proves $\mathsf{SparseIrred}_{\Z}\in\PP$.

Finally, a polynomial-time machine has only exponentially many branches of polynomial description length. A deterministic machine can enumerate them, simulate one at a time, and keep a polynomial-bit count. Hence $\PP\subseteq\PSPACE$, also directly. Theorem~\ref{thm:main} is proved.

%

%
\begin{remark}[Uniform gaps other than $1/32$]
The reduction only needs some fixed $\delta>0$, independent of the number field. Replacing $1/32$ by another fixed gap changes absolute constants and the fixed multiplier in $\log X$, not the polynomial-space or $\PP$ conclusions. Thus the logical statement ``uniform QRH implies sparse irreducibility in $\PP$'' is independent of the publication status of the particular zero-free theorem; its present unconditional application uses \cite{CuretonThorner2026}. A gap shrinking with $n$ would require a separate cutoff analysis and is not silently sufficient.
\end{remark}

\section{Conclusion}

We proved that irreducibility testing for sparse polynomials in $\Z[X]$
is coNP-hard under deterministic polynomial-time many-one reductions; equivalently, reducibility in $\Z[X]$ is NP-hard under the same
reductions; irreducibility is also NP-hard under polynomial-time Turing
reductions that use a single query. Hence, no polynomial-time
irreducibility test exists unless $\mathrm P=\mathrm{NP}$.
This settles the univariate integer case left open in
\cite{Karpinski1999CH,KaltofenKoiran2005,SparsityChallenges,Roche2018}; the corresponding questions over finite fields remain open.

\section*{Acknowledgment}
This paper is supported by the Strategic Priority Research Program of CAS Grants XDA0480502 and XDA0480503, NSFC Grants 12288201 and 92270001.

\section*{Declaration of AI Use}
The main results presented in this paper are obtained through an interactive collaboration between the authors and an artificial intelligence agent system, \textit{MechMath Agent Team (MMAT)}~\cite{MMAT} and GPT-Astra. 
The authors assume full responsibility for the paper’s content.

\IfFileExists{KLMM/klmm.bst}
  {\bibliographystyle{KLMM/klmm}}
  {\bibliographystyle{klmm}}
\bibliography{refs}

\clearpage
\appendix
\section{Lean Formalization Details}
\label{app:lean}

This appendix describes the formal development at the level of its
mathematical structure and its connection to the statements in the paper.
It covers the integer hardness proof and the integer $\PP$ and
$\PSPACE$ upper bounds. The complexity statements use complexitylib's
Turing-machine definitions and proved containments; the hardness proof
also uses its proved NP-completeness of exact-3-CNF satisfiability.
The integer upper bound retains the literature inputs specified below.

\subsection{Files and Formal Targets}

All project declarations below lie in the namespace \leanid{SparseIrred},
unless a further namespace is specified. Subdirectory paths below are
relative to \leanid{SparseIrred/}.
The module \leanid{SparseIrred.lean} imports the complete development.

\leanid{Construction/Defs.lean} defines the occurring-variable numbering,
the odd primes $q_i$, the literal and clause polynomials, and
$S,A,B,F$. The remaining files in \leanid{Construction/} prove the
Laurent-polynomial identities, the satisfying-assignment correspondence,
the algebraic-integer norm argument, and the cyclotomic factorization in
the satisfiable case.

The files \leanid{FFK/Dirichlet.lean}, \leanid{FFK/Lift.lean},
\leanid{FFK/Eisenstein.lean}, and \leanid{FFK/Ljunggren.lean} formalize
the simultaneous approximation, the lift to $\Z[x][y]$, Eisenstein's
criterion at $x-1$, and the Ljunggren argument for
Theorem~\ref{thm:gap}. The module \leanid{FFK/Main.lean} applies this
criterion to the formula-dependent construction.

\leanid{Sparse/Canonical.lean} defines canonical binary-sparse polynomial
codes and the irreducibility and reducibility languages.
The files in \leanid{Reduction/} define the total reduction and prove
its correctness on every source string. The files in \leanid{Machine/}
implement the sparse computation and prove its polynomial running time.
\leanid{Oracle/OneQuery.lean} adds exact query counting to the oracle
Turing-machine model and constructs the one-query reduction.

\leanid{Theorem8.lean} contains the four complexity conclusions; its name
reflects the numbering in an earlier set of progress notes.
\leanid{Irreducibility.lean} collects the main statements in the namespace
\leanid{SparseIrred.Irreducibility}. They include
\leanid{irreducibility_hardness}, \leanid{satisfiability_encoding},
\leanid{ljunggren_gapPoly}, \leanid{irreducible_gapPoly},
\leanid{exists_shifted_residue_modulus}, \leanid{lcm_primeCounting_bounds},
\leanid{prime_index_bounds}, and \leanid{polynomial_time_sparse_reduction}.

The integer upper-bound entry module is
\leanid{UpperBounds/Section3Complete.lean}. It collects the counting
argument, arbitrary real cutoffs, fixed positive zero-free gaps, and
the implemented machines in the namespace
\leanid{SparseIrred.UpperBounds}.

\subsection{Algebraic Construction and Gap Criterion}

The polynomial denoted by $F$ in the paper is
\leanid{Hpoly} for an arbitrary gap $M$, and \leanid{Hfinal} for the
chosen gap \leanid{Mval}. The statement \leanid{Irreducibility.satisfiability_encoding}
collects the nonvanishing, degree, reciprocity, rational-coprimality, and
root-of-unity assertions of Lemma~\ref{lem:carrier}.
The theorem \leanid{toLaurent_Apoly} proves that extracting $A$ as an
ordinary integer polynomial loses no Laurent terms. The Chinese
remainder step is \leanid{exists_root_realizing}; the equivalence
between satisfiability and a root of unity of $F$, for every $M$, is
\leanid{exists_rootOfUnity_root_iff}. Its converse uses the formalized
algebraic-integer norm argument in \leanid{Construction/Norm.lean}.

For Theorem~\ref{thm:gap}, the predicate \leanid{FFK.FFKThreshold}
is exactly the numerical inequality~\eqref{eq:gap-bound}.
The theorem \leanid{Irreducibility.exists_shifted_residue_modulus} gives the multiplier
$1\le d\le8^{|T|}$, the modulus $\kappa=\lfloor L/d\rfloor>2m$,
and the unique integer quotient with shifted residue strictly in
$(0,\kappa/2)$. The lifting identity is proved by
\leanid{FFK.kLift_mul_eq_of_mul_eq};
\leanid{FFK.irreducible_gapLift} supplies the Eisenstein step.
The palindromic identity and root-of-unity exclusion yield relative
primality of $F$ and its reciprocal, and mathlib's
\leanid{Polynomial.irreducible_of_mirror} completes the irreducibility
argument.

Internally, this argument uses mathlib's \leanid{Polynomial.mirror},
which equals \leanid{Polynomial.reverse} when the constant coefficient
is nonzero. The theorem \leanid{Irreducibility.ljunggren_gapPoly}
uses \leanid{reverse} directly and derives $K(0)\ne0$ from
$FF^*=KK^*$. Both parts of Theorem~\ref{thm:gap} accept coprimality in
$\Q[X]$; \leanid{Irreducibility.isRelPrime_of_rational_coprime} proves the
required relative primality in $\Z[X]$, also excluding integer
nonunit common factors by the constant coefficient $-1$ of $X^N-1$.

\subsection{Polynomial-Time Reduction and Complexity Conclusions}

The function \leanid{encPoly} represents an integer polynomial by its
nonzero coefficient--exponent pairs in increasing exponent order.
Coefficients have a sign and binary magnitude, and exponents are binary.
Theorems \leanid{encPoly_injective} and
\leanid{decodePoly_eq_some_iff} establish uniqueness and the exact
meaning of a canonical code. The languages \leanid{IrredLang} and
\leanid{RedLang} use mathlib's irreducibility predicate and factorization
into two nonunits, respectively. Noncodes belong to neither language;
on positive-degree polynomials, \leanid{reducible_iff_not_irreducible}
proves the complementary relation used by the reduction.

The function \leanid{reduce} handles malformed and degenerate inputs
as in the proof of Theorem~\ref{thm:main}. Its main branch applies
\leanid{encPoly} to \leanid{Hfinal} of the decoded formula $\Phi$.
The core theorem \leanid{Hfinal_irreducible_iff} proves, for every
nonempty CNF formula, that this polynomial is irreducible exactly when
the formula is unsatisfiable. The theorem \leanid{reduce_correct}
extends correctness to all binary source strings.

The prime bounds of Lemma~\ref{lem:prime} are established by two theorems:
\leanid{Machine.nth_prime_le} and \leanid{Machine.oddPrime_le}.
The two inequalities in~\eqref{eq:prime-bound} are combined in
\leanid{Irreducibility.lcm_primeCounting_bounds}.
The machine implementation computes only sparse term lists and binary
integers. Its Cobham-algebra membership and complexitylib's proved
Cobham characterization imply \leanid{reduce_mem_FP}.
The theorem \leanid{Irreducibility.polynomial_time_sparse_reduction} exhibits one deterministic machine
and one natural-coefficient polynomial bounding both running time and
output length, and verifies \eqref{eq:gap-bound} for \leanid{Mval}.

The four final statements in \leanid{Theorem8.lean} are
\begin{quote}
\leanid{irreducibility_coNPHard},\\
\leanid{reducibility_NPHard},\\
\leanid{irreducibility_NPHard_oneQuery},\\
\leanid{P_eq_NP_of_irreducibility_mem_P}.
\end{quote}
The oracle construction proves that its complete run makes exactly one
query. The strengthened statements
\leanid{P_eq_NP_of_irreducibility_test} and
\leanid{irreducibility_NPHard_oneQuery_of_codes} require correctness
only on canonical polynomial codes. No membership in NP or coNP is
asserted for the target languages.

\subsection{Integer Upper Bounds and Verified Machines}
\label{app:lean-integer-upper}

The number-field argument is formalized for the actual irreducible
factors of the input polynomial. It includes integral normalization of
nonmonic factors, discriminant bounds, the exceptional-prime bound, and
the correspondence between modular roots and degree-one prime ideals.
Thus Lemmas~\ref{lem:disc} and~\ref{lem:dedekind} are internal proofs,
not additional hypotheses. In the namespace
\leanid{SparseIrred.UpperBounds},
\leanid{integer_prime_root_average_concrete_real} proves the average
estimate of Proposition~\ref{prop:average} for every real $X>2$.
The cutoff analysis then establishes Corollary~\ref{cor:decision},
including inputs with repeated factors. The encoded decision procedure
also handles zero, units, constants, and nonprimitive polynomials.

The external analytic input is recorded by the named proposition
\leanid{EffectivePrimeIdealEstimate}, with a real-cutoff version
\leanid{EffectivePrimeIdealEstimateReal}. These state the effective
counting conclusion after supplying the uniform zero-free input of
\cite{CuretonThorner2026} to the estimate of
\cite{FreiSofosWidmer2026}. They quantify over number fields and cutoffs
before the input polynomial is chosen. The development proves the
choice of a single integer constant and all subsequent specializations.
It does not import the external zero-free project~\cite{AxiomMath2026}
or independently prove the analytic implication from zero-freeness to
the effective estimate. The other literature parameter is
\leanid{BinaryPrimalityInFP}, the binary-encoding form of the AKS
theorem~\cite{AgrawalKayalSaxena2004}.

The final integer theorem is
\begin{quote}
\leanid{section3_irreducibility_mem_PP_and_PSPACE_of_external_theorems}.
\end{quote}
It concludes membership of the actual encoded language
\leanid{SparseIrred.IrredLang} in complexitylib's \leanid{PP} and
\leanid{PSPACE}, with only the two literature inputs just described.
The string verifier, fair-bit guessing, preprocessing composition,
polynomial time bound on every branch, and exact acceptance probability
are proved in \leanid{UpperBounds/Machine/}. In particular, the
majority-machine argument is connected to a real machine, rather than
left as a counting specification or an assumed implementation
certificate. The theorem
\leanid{GeneralGap.section3_irreducibility_mem_PP_and_PSPACE_of_positive_gap}
extends the construction to every fixed positive gap, independent of
the input. The acceptance condition is strict majority, as required for
$\PP$; no bounded-error $\BPP$ claim is made.

The deterministic enumeration is also implemented. The theorem
\leanid{Machine.section3_deterministic_streaming_mem_PSPACE} applies to
the full encoded input and uses the same analytic and AKS inputs.
Separately, \leanid{Machine.section3_linear_streaming_core} proves an
affine space bound for every reachable configuration of the
trial-division core, without AKS, starting from its specified
preprocessed configuration. This latter result does not claim
linear-space preprocessing or an end-to-end linear-space machine on
arbitrary raw input.

\subsection{Axiom Boundary and Reproducibility}

The development uses no project-specific axioms and contains no
\leanid{sorry} or \leanid{admit} placeholders in the project proofs.
The repository-wide release audit checks 479 declarations and reports only
\begin{quote}
\leanid{propext},\quad \leanid{Classical.choice},\quad
\leanid{Quot.sound}.
\end{quote}
These are Lean's standard propositional-extensionality, classical-choice,
and quotient-soundness axioms. A separate traversal of the proof and type
dependencies of 463 entry points reaches 84,756 constants and no other
axioms. In particular, the source NP-hardness theorem, the machine
implementations, and the containment $\PP\subseteq\PSPACE$ are proved
dependencies. These checks certify the proof terms of the conditional
integer theorems; they do not prove their explicit literature
parameters.

For comparison with the earlier progress notes, the repository also
retains an alternative route with the general
Sawin--Shusterman--Stoll large-gap statement \leanid{SSSLargeGap} as an
explicit hypothesis. That statement is not a declared axiom and is not
used by the unconditional results of this paper. The FFK argument
needed here is proved directly.

The toolchain is fixed to \leanid{leanprover/lean4:v4.34.0-rc2}.
The dependency manifest \leanid{lake-manifest.json} fixes all package
revisions, including complexitylib at \leanid{edd0e9ec} and mathlib at
\leanid{e06eff5f}.
The 38 vendored sparse-arithmetic modules are listed in
\leanid{Vendor/MODULES.txt} and checked against the published
\leanid{Vendor/SHA256SUMS}; no sibling repository is needed. After
cloning \url{https://github.com/MechMath/sparse-irr}, the development
and audits can be checked from the repository root with
\begin{quote}
\begin{verbatim}
lake exe cache get
scripts/audit.sh
\end{verbatim}
\end{quote}
The audit script runs the complete build, the axiom and dependency
checks, and the vendored-source check, returning failure on an error.
The repository-wide release build completes 9,470 tasks.
Audit outputs are recorded in \leanid{docs/AXIOMS.txt} and
\leanid{docs/DEP_AUDIT.txt}. The full manuscript correspondence is in
\leanid{docs/PAPER_1021_MAP.md}; the section reports give more detailed
declaration maps and machine specifications. Historical module names
and comments may use earlier manuscript numbering. Background
discussion and illustrative examples, including Example~\ref{ex-1},
are not claimed to have separate Lean declarations.

\end{document}